\documentclass[prb,floatfix,twocolumn,showpacs,amsmath,amssymb]{revtex4}
\usepackage{graphicx,color}% Include figure files
\usepackage{dcolumn}% Align table columns on decimal point
\usepackage{bm}% bold math
\begin{document}

\title{Theory of spin density profile in
  the magnetization plateaus of ${\rm SrCu_2(BO_3)_2}$}
\author{Shin Miyahara,$^{1}$ Federico Becca,$^{1,2}$ and Fr\'ed\'eric Mila$^{1}$}
\affiliation{
${^1}$ Institut de Physique Th\'eorique, Universit\'e de Lausanne, CH-1015 Lausanne, Switzerland \\
${^2}$ INFM-Democritos, National Simulation Centre, and SISSA I-34014 Trieste, Italy.
}

\date{\today}

\begin{abstract}
The two-dimensional spin-gap system ${\rm SrCu_2(BO_3)_2}$ shows unique 
physical properties due to the low-dimensionality character and the strong
quantum fluctuations.
Experimentally, $1/8$-, $1/4$-, and $1/3$-plateaus have been observed 
in the magnetization curve under magnetic fields up to $70$ Tesla, and
the $1/2$-plateau is expected to be stabilized at higher magnetic fields. 
We argue that spin-lattice effects are necessary to describe the 
superstructures at the plateaus, and we propose a simple microscopic 
model of spins interacting adiabatically with the lattice to
reproduce the main features of the recent experimental results by nuclear 
magnetic resonance.
\end{abstract}
%%%%%%%%%%%%%%%%%%

\maketitle

\section{introduction}

\begin{figure}
\vspace{2mm} 
\includegraphics[width=0.45\textwidth]{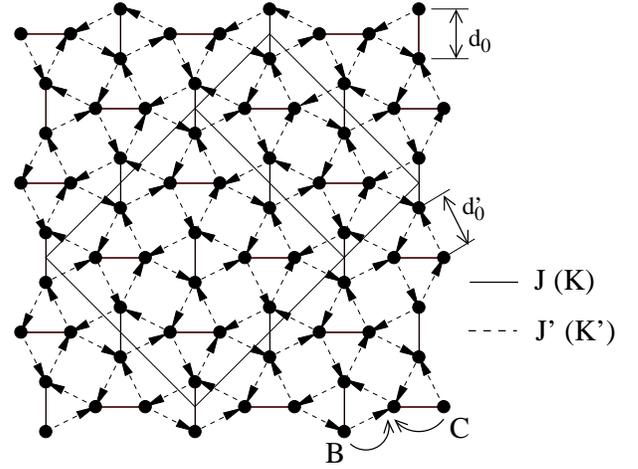}
\caption{\label{fig:2d-model}
The orthogonal dimer 
lattice. The continuous and dashed lines indicate
the antiferromagnetic interactions and the elastic couplings for 
nearest neighbor and next-nearest neighbor sites, respectively.
The nearest neighbor and the next-nearest neighbor equilibrium distances
are indicated by $d_0$ and $d_0^\prime$, respectively.
The 16-site square cluster and 24-site rectangular cluster used in
the calculations are shown by thin solid lines.
The transferred hyperfine couplings $B$ and $C$ 
(see Section~\ref{section:comp}) are also shown.
The arrows define the direction from the site $i$ to the site $j$
for the Dzyaloshinsky-Moriya interaction (see Section~\ref{section:DM}).
}
\end{figure}

The possibility to obtain low-dimensional quantum spin systems that
do not order magnetically at very low temperature, or even down to
zero temperature, is currently a subject of great interest. 
Although there are many examples of one-dimensional (1D) or quasi-1D
systems such as ${\rm SrCu_2O_3}$ ($S=1/2$ ladder),~\cite{ladder}
${\rm Y_2BaNiO_5}$ (Haldane chain),~\cite{haldane}
${\rm CuGeO_3}$ and ${\rm LiV_2O_5}$ (prototype of frustrated $S=1/2$ 
chains),~\cite{chain1,chain2}
for some time, the only example of a  two-dimensional (2D)
system with a singlet ground state and a finite gap
to magnetic excitations was the vanadium oxide
${\rm CaV_4O_9}$ (Ref.~\onlinecite{taniguchi95,ueda96,troyer96}).
Therefore, the recent discovery of ${\rm SrCu_2(BO_3)_2}$ 
(Ref.~\onlinecite{kageyama99}) represents a breakthrough in this direction
and paves the way for further research on this unconventional state of
matter. This compound has a layered structure where 
stacking layers of ${\rm CuBO_3}$ are
intercalated by magnetically inert layers of ${\rm Sr}$.
A spin $S=1/2$ resides on each ${\rm Cu^{2+}}$ ion, forming a 2D
orthogonal dimer lattice~\cite{kageyama99} 
(see Fig.~\ref{fig:2d-model}).
It turns out that the magnetic properties of 
${\rm SrCu_2(BO_3)_2}$ are very well described by the
2D orthogonal dimer Heisenberg model,~\cite{miyahara99} 
which is topologically equivalent to the 
2D Shastry-Sutherland model:~\cite{shastry81}
\begin{equation}\label{eq:model} 
  {\cal H} =J \sum_{\rm (n.n.)} {\bf S}_i \cdot {\bf S}_j
  +J^{\prime} \sum_{\rm (n.n.n.)} {\bf S}_i \cdot {\bf S}_j,
\end{equation}  
here ${\bf S}_i=(S_i^x, S_i^y, S_i^z)$ is the spin-$1/2$ operator at the
site $i$ and the notations ${\rm (n.n.)}$ and ${\rm (n.n.n.)}$ stand for 
nearest neighbor and next-nearest neighbor sites, respectively.
In the parameter range $J^{\prime}/J < 0.68$ (Ref.~\onlinecite{koga00})
the ground state is exactly known
to be the product of dimer singlets:~\cite{shastry81}
\begin{equation}\label{eq:dimer-singlet}
  |\Psi\rangle  
  = \prod_a \frac{1}{\sqrt{2}}(| \uparrow \downarrow \rangle_a -
  | \downarrow \uparrow \rangle_a ),
\end{equation}
where $a$ indicates the dimer bond connected by the superexchange coupling $J$.
In Ref.~\onlinecite{miyahara00b}, it has been shown that the values of the
antiferromagnetic superexchange are such that $J^{\prime}/J = 0.635$,
with $J = 85$ Kelvin, indicating that ${\rm SrCu_2(BO_3)_2}$ can be 
described by a 2D spin system whose ground state is exactly known.

The very particular nature of the ground state induces unique features
in the spin excitations at low temperatures. First of all,
there is a finite gap to the magnetic excitations. 
This spin gap has been observed in several experiments
and was estimated to be about
$35$ Kelvin.~\cite{nojiri99,kageyama00c,lemmens00b,room00} 
Moreover, the triplet excitations have an almost localized 
nature because of the orthogonality of the $J$ bonds.~\cite{miyahara99}
Such a behavior was revealed experimentally, 
using inelastic neutron scattering,~\cite{kageyama00c} as an almost flat
triplet dispersion.

The localized nature of the triplet excitation leads also to 
the most spectacular phenomenon of this system: 
When an external magnetic field, up to $70$ Tesla, is applied, the system
shows magnetization plateaus
corresponding to $1/8$, $1/4$, and $1/3$ of the full ${\rm Cu^{2+}}$ 
moment (another plateau at $1/2$ of the full ${\rm Cu^{2+}}$ moment is
likely to be stabilized at even higher magnetic 
fields).~\cite{kageyama99,onizuka00}
By applying an external magnetic field, the density of
the triplets can be tuned and it is found that the magnetization stays 
constant for particular ranges of the external field.
It is expected that the plateaus are due to a crystallization
of the triplets for particular commensurate values of the 
magnetization and are due to the small ratio between the kinetic
and the interaction energy of the triplets.
To our knowledge, ${\rm SrCu_2(BO_3)_2}$ is the first example 
of a 2D quantum spin system which shows magnetization plateaus.

So far, the magnetization plateaus have been studied mostly
in 1D systems.~\cite{hida94,okamoto95,tonegawa96,narumi98} 
In particular, Yamanaka, Oshikawa, and Affleck~\cite{oshikawa97} have found 
a simple necessary condition for the existence 
of plateaus in 1D systems.
Denoting by $l$ the period of the ground state in the presence of the
external field, by $S$ the magnitude of the spin and by $m$ the 
magnetization per site (in unit of $g\mu_B$), the occurrence of a 
magnetization plateau is only possible when the condition 
\begin{equation}\label{eq:plateau_condition}
  l(S-m) = {\rm integer},
\end{equation}
is satisfied.
It is important to mention that $l$ can be different from the period of the 
lattice. The first example of a magnetization plateau accompanied by
such a symmetry breaking is a $S=1/2$ Heisenberg chain 
with next-nearest-neighbor
and alternating nearest-neighbor interactions.~\cite{tonegawa98,totsuka98}

Recently, Oshikawa extended the Lieb-Schultz-Mattis argument to
quantum many-particle systems with a conserved particle number on 
a periodic lattice in arbitrary dimensions and showed
that the condition of Eq.~(\ref{eq:plateau_condition}) for the plateau 
is still valid in arbitrary dimensions.~\cite{oshikawa00}
Indeed, several theoretical works concerning 
the plateaus for ${\rm SrCu_2(BO_3)_2}$
indicate that the criterion of Eq.~(\ref{eq:plateau_condition}) is satisfied
and, except for the plateau at $1/2$, the ground states at the plateaus
are accompanied by a breaking of the translational 
symmetry.~\cite{momoi00,momoi00b,miyahara00,fukumoto00,fukumoto01,misguich01}
 
Recently, Kodama and collaborators~\cite{kodama02b} performed a nuclear 
magnetic resonance (NMR) measurement at the $1/8$-plateau, 
for an external magnetic field 
$H=27.6$ Tesla and a temperature of $35$ mK, and observed the presence 
of at least $11$ {\it different} sites (i.e., different values of the local 
magnetization), indicating a clear breaking of the translational symmetry.
So far, from a theoretical point of view, the superstructures at the 
plateaus have been studied only by an effective hard-core boson model.
In this approximation, the triplet with $S^z = 1$ is represented
by a hard-core boson and the dimer singlet by a vacancy. An effective 
Hamiltonian is derived by perturbation theory and the magnetization curve
and superstructures have been calculated by solving 
it.~\cite{momoi00,momoi00b,miyahara00,fukumoto00,fukumoto01}
In this picture, two different unit cells with $16$ sites
have been proposed to describe this plateau:
i) a $16$-site square unit cell and ii) a rhomboid unit 
cell.~\cite{miyahara00}
However, the ground state at the $1/8$-plateau is described by a state
where one of the eight singlets is promoted to a triplet within the
unit cell. Thus there are only two different sites,
corresponding to the singlet and the triplet states, and it cannot 
reproduce the very rich texture of the magnetization observed 
in NMR experiments. 
Therefore, in order to reproduce the experimental data, the simple
hard-core boson model is not sufficient and it is necessary to
consider the original spin Hamiltonian.

From the results of the hard-core boson calculations,
as well as from the result of Ref.~\onlinecite{oshikawa00},
it is expected that the ground states at plateaus are degenerate.
On the other hand, a magnetization structure will only be observed if an extra 
mechanism selects one of the ground states, since linear combinations of the 
ground states can lead to an {\it arbitrary} magnetization. 
For instance, a uniform linear combination would lead to a uniform 
magnetization. In the real material, this mechanism could be due to 
pinning by impurities, or to a lattice distortion. 
Actually, a strong motivation for considering lattice effects in
${\rm SrCu_2(BO_3)_2}$ is given by the pronounced softening of the sound 
velocity observed at the edges of the magnetization 
plateau.~\cite{zherlitsyn00,wolf01}
On the theoretical side, one could try to select a magnetization texture
by imposing an external, very small symmetry breaking field, 
as is often done for instance for the dimerized state
in a spin-Peierls system. However, in the present case, we do 
not know a priori the magnetization texture, and imposing a specific field 
would bias the results. Therefore, we prefer the more physical way 
which consists in coupling the system to phonons.

The paper is organized as follows. In Section~\ref{section:method}, 
we introduce the spin-phonon Hamiltonian and explain the method. 
In Section~\ref{section:results}, we present the 
results for the superstructures at $1/8$-, $1/4$-, $1/3$- and $1/2$-plateaus.
In Section~\ref{section:comp} we compare our theoretical results for 
the $1/8$-plateau to the experimental ones, 
including in addition the effects of inter-layer coupling 
and Dzyaloshinsky-Moriya interaction.
Finally, Section~\ref{section:conclusions} is devoted to the conclusions 
and the discussion.

\section{Model and method}\label{section:method}

In this Section we introduce the spin-phonon Hamiltonian and we describe
the method that we use to characterize the spin texture of the 
different magnetization plateaus.
We consider the $S=1/2$ orthogonal dimer model coupled to 
adiabatic phonons that we study by exact diagonalizations of
finite clusters with a self-consistent Lanczos algorithm. 
In this approach, the adiabatic phonons are described by classical 
variables, related to the displacements of the lattice sites.
The full spin-phonon Hamiltonian is defined on a 2D 
orthogonal dimer lattice of 
$N$ sites by:
\begin{eqnarray}
&&{\cal H} = 
\sum_{\rm (n.n.)} \left \{ 
J(d_{ij}) {\bf S}_i \cdot {\bf S}_j +
\frac{K}{2} \left (
\frac{\|\delta {\bf r}_i-\delta {\bf r}_j\|}{d_{ij}^{0}} \right )^2
\right \} + \nonumber \\
&&\sum_{\rm (n.n.n.)} \left \{ 
J^{\prime}(d_{ij}) {\bf S}_i \cdot {\bf S}_j +
\frac{K^{\prime}}{2} \left (
\frac{\|\delta {\bf r}_i-\delta {\bf r}_j\|}{d_{ij}^{0}} \right )^2
\right \},
\label{hamilt}
\end{eqnarray}
here $J(d_{ij})$ and $J^{\prime}(d_{ij})$ are the antiferromagnetic 
superexchange couplings, which depend on the relative distance 
$d_{ij}=\|{\bf R}_{i}^{0}+\delta {\bf r}_i-{\bf R}_{j}^{0}-\delta {\bf r}_j\|$
between sites 
$i$ and $j$. $K$ and $K^{\prime}$ are the elastic coupling constants,
$\delta d_{ij}=d_{ij}-d_{ij}^{0}$, and 
$d_{ij}^{0}=\|{\bf R}_{i}^{0}-{\bf R}_{j}^{0}\|$ is the
equilibrium distances between Copper sites.

For small displacements of the ${\rm Cu}$ sites, it is possible to linearize 
the antiferromagnetic couplings around their equilibrium values,
$\delta d_{ij} \simeq ({\bf R}_{i}^{0}-{\bf R}_{j}^{0}) \cdot 
(\delta {\bf r}_{i}-\delta {\bf r}_{j})$, and therefore we expect
that in general:~\cite{harrison}
\begin{equation}
\label{jj}
J(d_{ij})=J \left ( \frac{d_{ij}^0}{d_{ij}} \right )^\alpha \simeq
J \left ( 1 - \alpha \frac{\delta d_{ij}}{d_{ij}^{0}} \right ).
\end{equation}
In the following, we will denote by $\alpha$ and $\alpha^{\prime}$ the two 
(in principle different) exponents for $J(d_{ij})$ and
$J^{\prime}(d_{ij})$, respectively.

With a self-consistent Lanczos diagonalization of finite clusters, one can find
the optimal configuration of the bond lengths and local spin configurations
for given values of the coupling parameters 
($\alpha$, $\alpha^{\prime}$, $K$ and $K^{\prime}$) 
and given total spin $S$ quite easily:
Starting from a random choice of the atomic displacements, one 
just has to improve iteratively the total energy by
changing the lattice parameters until a stationary
configuration is reached. In all cases, we have verified that 
this configuration is stable
and unique (up to trivial symmetry operations) by starting from
different initial distributions, so it must be the global minimum.
This method has been already used for other 1D, quasi-1D and 2D 
spin systems interacting adiabatically with the lattice.~\cite{poilblanc,becca}
With respect to other approximate approaches, this method has the great 
advantage that it gives unbiased results even for strongly frustrated
systems, where other numerical methods can be highly questionable.
It is worth noting that, in our self-consistent Lanczos method, the only
approximation is to consider the adiabatic limit for the phonons.
Once we restrict our calculation to this case, we obtain exact results
for the lattice distortions on the chosen finite cluster.

In the following, we consider unit cells containing $N=16$ sites or, in
some case, up to $N=24$ sites with periodic boundary conditions 
(see Fig.~\ref{fig:2d-model}). 
In our simple microscopic model, we assume that the magnetic field directly
couples to the total spin of the system, stabilizing the states with higher
total spin. Because of the finiteness of our cluster,
in order to study the plateau at $1/n$, we assume that 
the external field is able to stabilize a state with a given magnetization
and we fix the total spin of the cluster 
(by fixing $S^z_{tot}=\sum_i S^z_i=\frac{N}{2n}$).
It is worth mentioning that, within our self-consistent Lanczos method, 
we are not able to 
study the actual width of the magnetization plateau in the thermodynamic limit.
Nonetheless, once the existence of a given plateau is assumed, we can produce 
important insight into the local structure of the spins inside the cell.

In the following calculations, we consider the case of 
$\alpha^\prime = 1.75 \alpha$ and $K = K^\prime = 750 J^{\prime}$
but we want to stress that all the results are not
specific of this particular choice and, qualitatively, 
we found similar results also for different values of these parameters.
The parameters $K$ and $K^\prime$ are only effective 
parameters that cannot be directly matched to the phonon dispersion. 
An appropriate model of elastic constants should include the springs
between all nearest neighbor atoms, but this does not improve the
calculation because the dependence of the spring constants
on the actual position of all atoms is
not exactly known. Nevertheless, the order of magnitude of the elastic 
constants is expected to be similar in all oxides.
For instance, for ${\rm CuGeO_3}$, in order to have relative displacements
smaller than a percent -- as observed experimentally in the spin-Peierls 
phase -- the values of the elastic constants are required to be of the 
order of $100000$ Kelvin.~\cite{becca2} Finally, superexchange theory
suggests that typical value for $\alpha$ in oxides are in the
range $6$-$14$ (Ref.~\onlinecite{harrison}).

The bond length of nearest neighbor bond $d_0$ is taken as the unit of the
length, which is $2.91 \AA$ at $100$ Kelvin,
and the length of next-nearest neighbor bond $d_0^\prime$
is assumed to be given by $d_0^\prime = 1.75 d_0$
(Ref.~\onlinecite{sparta01}).
Finally, we note that the Hamiltonian~(\ref{hamilt}) is invariant under the 
rescaling
$\alpha \to \lambda \alpha$, $\alpha^\prime \to \lambda \alpha^\prime$,
$K \to \lambda^2 K$, $K^\prime \to \lambda^2 K^\prime$, and
$\delta {\bf r}_{i} \to \delta {\bf r}_{i}/\lambda$, where $\lambda$ is 
the rescaling parameter.
This allows us to fix one parameter among $\alpha$, $\alpha^\prime$, 
$K$ and $K^\prime$. Note that the physical values of the magnetic couplings
are unaffected by this transformation.

\section{Effect of spin-phonon coupling at plateaus}\label{section:results}

\subsection{$1/8$-plateau}\label{section:1/8-plateau}

\begin{figure}
\vspace{2mm} 
\includegraphics[width=0.40\textwidth]{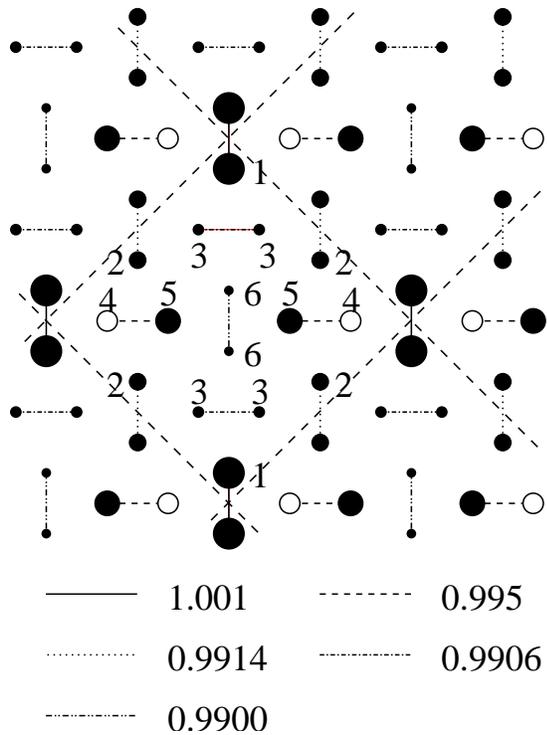}
\caption{\label{fig:plateau0125a}
Spin density profile for $1/8$-plateau for the $16$-site square cluster. 
Full (empty) circles indicate sites with magnetization along (opposite to) 
the external field and the size of the circles is proportional to the
spin amplitude. The bond lengths for $\alpha = 10$ are also shown.
}
\end{figure}

\begin{figure}
\vspace{2mm} 
\includegraphics[width=0.40\textwidth]{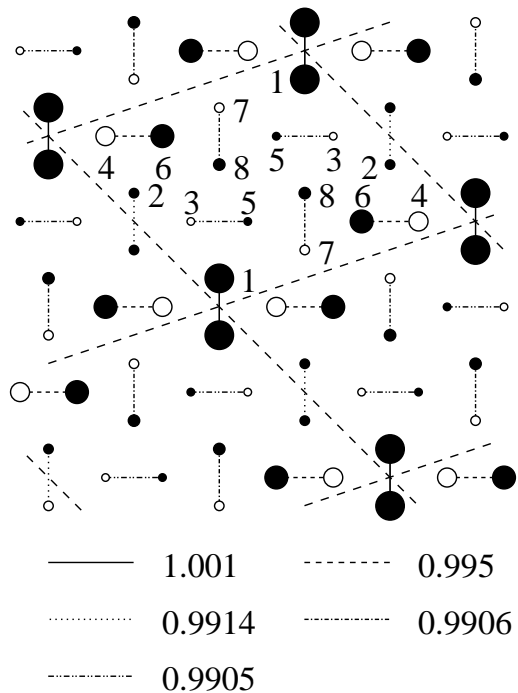}
\caption{\label{fig:plateau0125b}
Spin density profile for $1/8$-plateau for the rhomboid cluster of $16$ sites. 
Full (empty) circles indicate sites with magnetization along (opposite to) 
the external field and the size of the circles is proportional to the
spin amplitude. The bond lengths for $\alpha = 10$ are also shown.
}
\end{figure}

In order to describe the two most probable triplet patterns that were
previously suggested for the $1/8$-plateau by the hard-core boson
approach,~\cite{miyahara00} we have 
diagonalized the Hamiltonian~(\ref{hamilt}) on two $16$-site clusters
corresponding to different unit cells, 
the $16$-site square cluster and the rhomboid cluster in the sector
with $S^z_{tot}=1$ (see Figs.~\ref{fig:plateau0125a} 
and~\ref{fig:plateau0125b}).
The results for the two cases turn out to be qualitatively different:
In the square cluster we find that the ground state has only six different 
sites, that is six different values of the local magnetization 
$\langle S_i^z \rangle$ (see Fig.~\ref{fig:plateau0125a}), whereas the 
rhomboid cluster contains eight different sites with different local 
magnetization (see Fig.~\ref{fig:plateau0125b}).
The energy difference between these two states is very small (of the 
order of $10^{-5} \div 10^{-6} J$ per site).
In both cases, the magnetization is centered around one strongly
polarized dimer, that, in the following, we will denote by ``triplet'', 
with Friedel-like oscillations in the spin amplitude.
It is worth noting that we obtain both positive and negative magnetizations,
and in particular, in both clusters, there is a large negative spin, 
just near to the strongly polarized dimer. 

One of the main features of these results is the existence of two sites 
with large and positive polarization and one site with large and negative 
polarization, which is in agreement with the NMR experimental 
finding.~\cite{kodama02b}
The Friedel-like oscillations decay quite fast in space, and far from the 
``triplet'' dimer the local magnetization is very small. 
For this reason, we expect rather small size effects for our finite cluster
calculation and we believe that even the $16$-site lattice can represent
quite well the real system. Of course, we cannot rule out that a 
larger number of different sites (with very small magnetizations) exists when a 
bigger cluster is considered, but the very fast decay in the 
oscillation of the local magnetization clearly indicates
that our small $16$-site cluster is able to capture the main ingredients
of the true ground state: Two sites with large and positive 
$\langle S_i^z \rangle$, one site with large and negative
$\langle S_i^z \rangle$ and a bunch of sites with rather small (positive
and negative) magnetizations.
Unfortunately, the fact that we do not want to impose any external
constraint on the spin pattern prevents us
from using the lattice symmetries in the 
Lanczos diagonalization, and, in order to obtain the optimal
lattice (and spin) configuration, it is necessary to perform the
diagonalization several times up to convergence. 
These two facts make the calculation very heavy, and the next cluster,
with $32$ sites, which is consistent with the $1/8$-plateau, is beyond
the present computational possibilities.

\begin{figure}
\vspace{2mm} 
\includegraphics[width=0.40\textwidth]{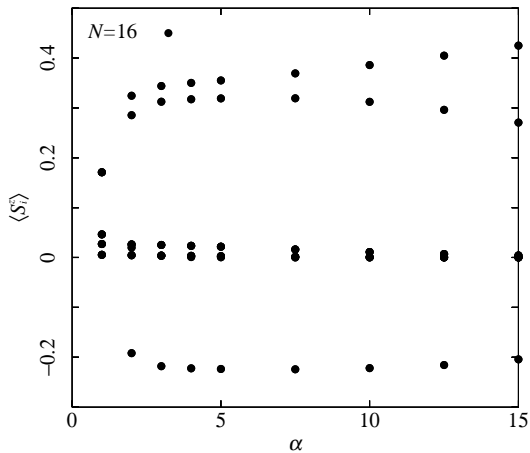}
\caption{\label{fig:sz0125a}
Local magnetization for the six different sites of the $16$-site
square cluster for the $1/8$-plateau as function of the spin-phonon coupling
$\alpha$.
}
\end{figure}

\begin{figure}
\vspace{2mm} 
\includegraphics[width=0.40\textwidth]{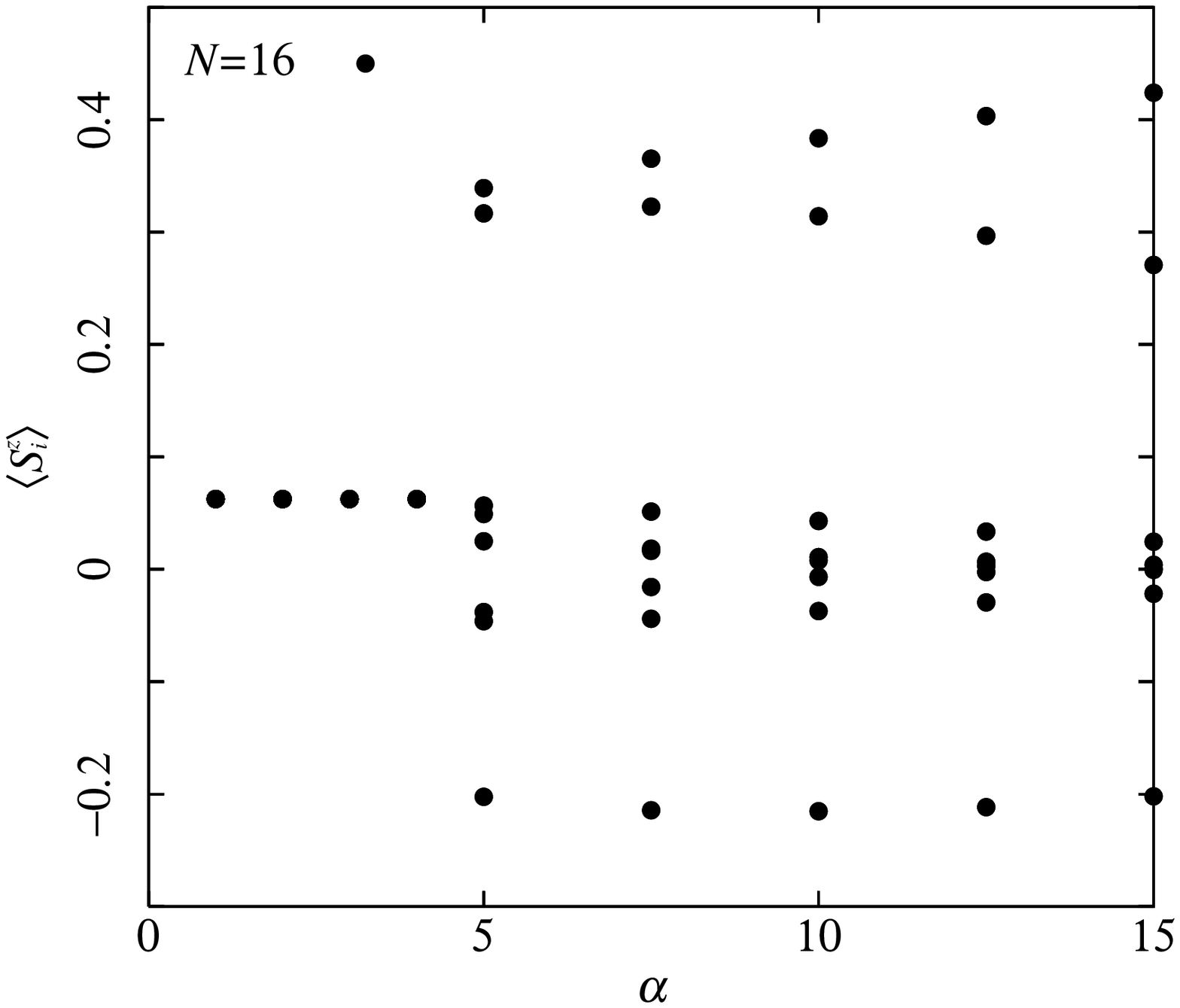}
\caption{\label{fig:sz0125b}
Local magnetization for the eight different sites of the rhomboid cluster 
for the $1/8$-plateau as a function of the spin-phonon coupling $\alpha$.
}
\end{figure}

In the optimized lattice configuration, all the translational symmetries are 
broken. The ground state is eight-fold degenerate for the 
square unit cell. On the other hand, in the rhomboid case,
two types of unit cell are possible and so the ground state
is sixteen-fold degenerate.
The behavior of the local magnetization as a function of the spin-phonon
coupling is reported in Figs.~\ref{fig:sz0125a}
and~\ref{fig:sz0125b} for the square and the rhomboid cluster,
respectively.
Notice that below a critical value of the spin-phonon coupling
$\alpha$, a state with a different number of sites is stabilized (four
sites for the square cluster and only one for the rhomboid cluster).
This is due to the fact that in the 16-site square (rhomboid) cluster
without spin-phonon coupling, the ground state is four-fold (two-fold) 
degenerate, and the other low-lying states are separated by a 
finite-size gap. 
Thus for small enough spin-phonon coupling, the ground state 
corresponds to a linear combination of only four (two) states, and
it is only possible to mix all the eight low-lying states with a sufficiently
large coupling, resulting into a state with six (eight) different sites.
The critical value of the spin-phonon coupling is much larger for
the rhomboid cell (of the order of $5$) than for the square unit cell
(between $1$ and $2$), see Figs.~\ref{fig:sz0125a} and~\ref{fig:sz0125b}.
As in the spin-Peierls case,~\cite{becca2} 
it is expected to decrease by enlarging the lattice size; unfortunately, for
reasons given just above, it is impossible to attempt a size scaling
of this value.

For both unit cells, we recover the hard-core boson 
results of Ref.~\onlinecite{miyahara00} in the extreme limit of infinite 
spin-phonon coupling ($\alpha \to \infty$), where all the magnetization 
is carried by a localized triplet dimer, and all the other dimers are 
perfect singlets.

Finally, we want to make a remark on the lattice displacements. 
The magnetic energy gain is related to a shrinking of the bond length of 
the ``singlets'', whereas the ``triplet'' dimers enlarge their bond
length. The bond lengths for $\alpha = 10$ for square (rhomboid) unit
cell are shown in Fig.~\ref{fig:plateau0125a} (Fig.~\ref{fig:plateau0125b}).
However, for realistic values of the elastic constants and the spin-phonon 
coupling, these displacements are very small 
-- of the order of one percent or less -- and 
we do not think that there is any chance at the moment to detect such a tiny 
structural distortion with X-Rays in a field of $27.6$ Tesla.

\subsection{$1/4$-plateau}

\begin{figure}
\vspace{2mm} 
\includegraphics[width=0.40\textwidth]{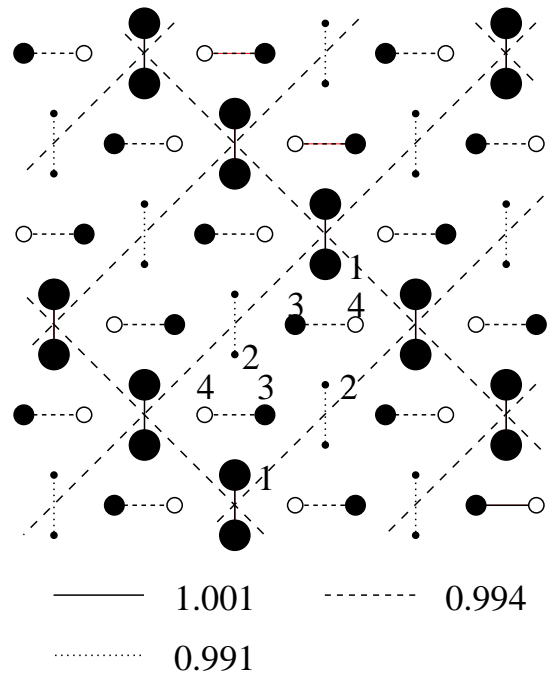}
\caption{\label{fig:plateau025}
Spin density profile for $1/4$-plateau. Full (empty) circles indicate 
sites with magnetization along (opposite to) the external field and 
the size of the circles is proportional to the spin amplitude.
The bond lengths for $\alpha = 10$ are also shown.
}
\end{figure}

Experimentally the next magnetization plateau is at $1/4$ of the total
${\rm Cu^{2+}}$ moment. From the hard-core boson 
calculations,~\cite{miyahara00} it comes out that the unit cell 
is just half of the previous $16$-site square cluster. 
Therefore, for this plateau, we consider 
the $16$-site square lattice and we perform the self-consistent Lanczos 
method in the sector $S^z_{tot}=2$. Moreover, we also report some results
for a larger $24$-site rectangular lattice with $S^z_{tot}=3$.

\begin{figure}
\vspace{2mm} 
\includegraphics[width=0.40\textwidth]{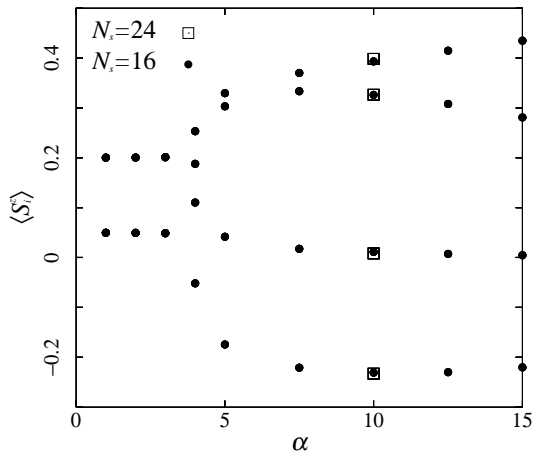}
\caption{\label{fig:sz025}
Local magnetization for the four different sites for the $1/4$-plateau as
a function of the spin-phonon coupling $\alpha$.
The full dots indicate the results for the $16$-site cluster 
and the empty squares the results for the $24$-site one
at $\alpha = 10$
}
\end{figure}

The typical outcome of our exact calculation is shown in 
Fig.~\ref{fig:plateau025}: We find four different sites, three with a positive
and one with a negative magnetization. Also in this case, the 
``triplet'' dimer is almost localized and the system shows
Friedel-like oscillations on the spin amplitude, the nearest site to
the ``triplet'' having negative $\langle S^z \rangle$.
The ground state is eight-fold degenerate and the eight states are found to
be connected by simple symmetry operations of the lattice, like 
translations and/or reflections.
The behavior of the local magnetization as a function of the spin-phonon
coupling is reported in Fig.~\ref{fig:sz025}
and it clearly indicates that by increasing $\alpha$, the total
magnetization concentrates progressively in the ``triplet'' dimer and the
local magnetizations of the other dimers tend to zero.
In the extreme limit of $\alpha \to \infty$, the picture corresponds to
the hard-core boson approximation, where there is only one triplet
that carries all the magnetization.~\cite{miyahara00,fukumoto01}

Notice that the spin texture is built up by diagonal stripes of strongly
polarized dimers, i.e., ``triplets'', intercalated by dimers with a small
polarization. This picture comes out from the fact that there are
repulsive interactions between the triplets,
which make the stripe structure stable
(this can be seen for example in the perturbation 
theory~\cite{miyahara00,fukumoto01}).

\subsection{$1/3$-plateau}

\begin{figure}
\vspace{2mm} 
\includegraphics[width=0.40\textwidth]{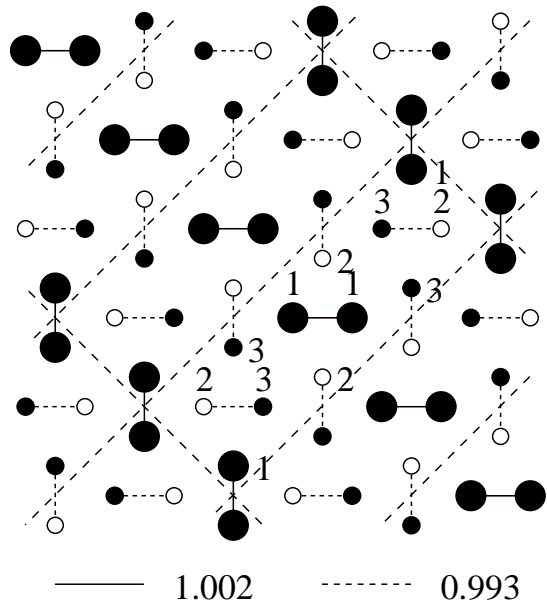}
\caption{\label{fig:plateau033}
Spin density profile for $1/3$-plateau. Full (empty) circles indicate 
sites with
magnetization along (opposite to) the external field and the size of the 
circles is proportional to the spin amplitude.
The bond lengths for $\alpha = 10$ are also shown.
}
\end{figure}

\begin{figure}
\vspace{2mm} 
\includegraphics[width=0.40\textwidth]{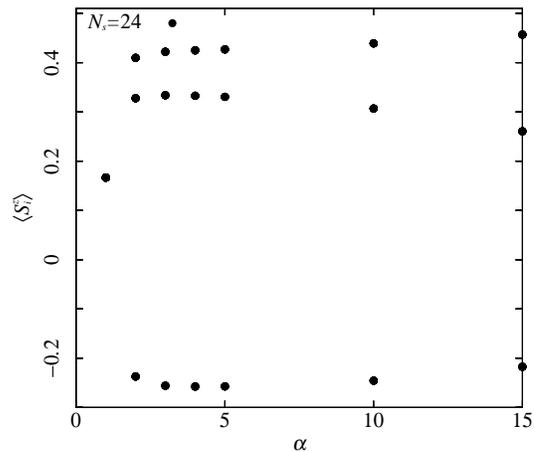}
\caption{\label{fig:sz033}
Local magnetization for the three different sites for the $1/3$-plateau as
a function of the spin-phonon coupling $\alpha$.
}
\end{figure}

In order to study the $1/3$-plateau, we are forced to consider the 
$24$-site cluster with $S^z_{tot}=4$ because in the $16$-site cluster
the corresponding spin sector is not present.
It is worth noting that this cluster breaks the reflection symmetries
of the original orthogonal dimer lattice; on the other hand, the shape
of this cluster fits well the pattern suggested in previous hard-core boson
calculations.~\cite{momoi00,momoi00b,miyahara00,fukumoto00}
For this plateau, the Lanczos results display three different values
of the local spin (see Fig.~\ref{fig:plateau033}): A large and positive
site, which corresponds to the strongly polarized ``triplet'' dimer, 
and two small sites (one positive and one negative), building up the
other ``singlet'' dimers.
The local value of the magnetization as a function of the spin-phonon
coupling is reported in Fig.~\ref{fig:sz033}.
The spin texture has a rectangular shape, contains $12$ sites, and it is
twelve-fold degenerate.

As for the other cases, the Lanczos
results for infinite spin-phonon coupling coincide with the hard-core
boson ones. As in the case of the $1/4$-plateau, the spin texture
configuration shows diagonal ``triplet'' stripes, and this configuration
is stabilized by the repulsion between strongly polarized 
dimers.~\cite{momoi00,momoi00b,miyahara00,fukumoto00}

\subsection{$1/2$-plateau}

\begin{figure}
\vspace{2mm} 
\includegraphics[width=0.40\textwidth]{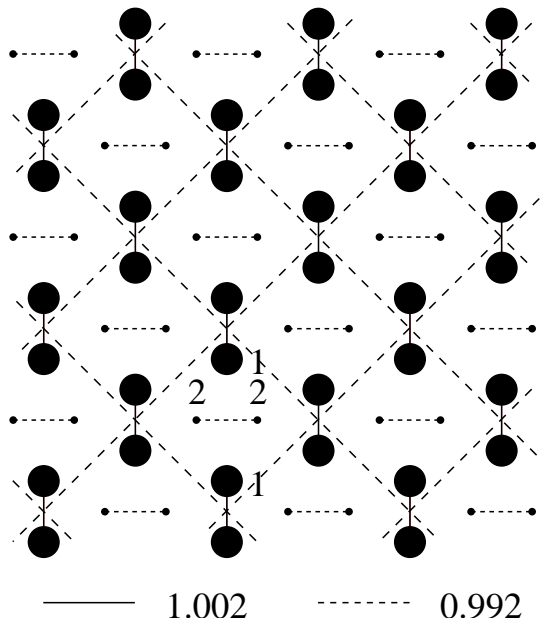}
\caption{\label{fig:plateau05}
Spin density profile for $1/2$-plateau. Full circles indicate sites with
magnetization along the external field and the size of the circles is 
proportional to the spin amplitude.
The bond lengths for $\alpha = 10$ are also shown.
}
\end{figure}

\begin{figure}
\vspace{2mm} 
\includegraphics[width=0.40\textwidth]{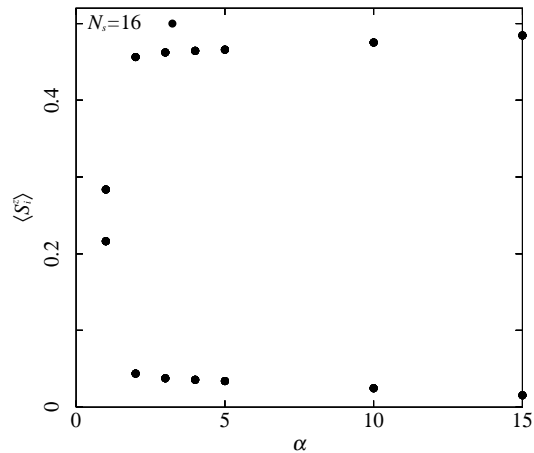}
\caption{\label{fig:sz05}
Local magnetization for the two different sites for the $1/2$-plateau as
a function of the spin-phonon coupling $\alpha$.
}
\end{figure}

For completeness, we finally consider the calculations 
for the $1/2$-plateau, by taking the $16$-site cluster with
$S^z_{tot}=4$. The typical spin configuration is shown in
Fig.~\ref{fig:plateau05}; here, we have only two different sites, both
with positive magnetization, corresponding to two different dimers: 
A ``triplet'', strongly polarized, and a ``singlet'', with a 
small magnetization.
Hence, we obtain a square unit cell that contains only four sites.
The evolution of the two local polarizations as a function of the spin-phonon
coupling is reported in Fig.~\ref{fig:sz05}.
In this case, only the symmetry which interchange the
``triplet'' with the ``singlet'' is broken, whereas the translational 
symmetry is preserved. Therefore, the ground state is
only two-fold degenerate, corresponding to the two possible choices of
the ``triplet'' position in the unit cell.

\section{Comparison with NMR experiments at 1/8-plateau}\label{section:comp}

As stated previously, at present, due to the very high magnetic field and 
the low temperature needed to stabilize the plateaus, 
it is only possible to perform accurate NMR experiments
for the $1/8$-plateau, which can give useful insight 
into the local spin texture.~\cite{kodama02b}
Therefore, in the following, we make a more detailed analysis of our
numerical results for this magnetization plateau.

\begin{figure}
\vspace{2mm} 
\includegraphics[width=0.4\textwidth]{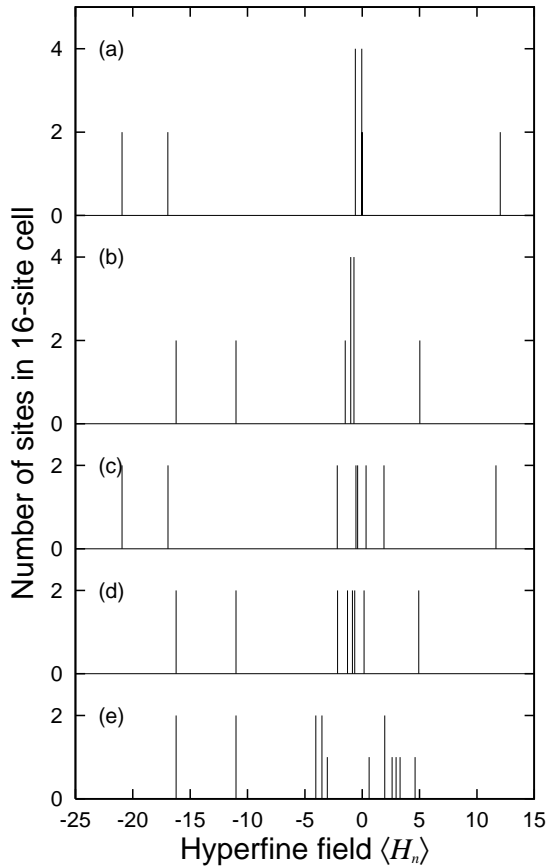}
\caption{\label{fig:comp_exp}
  Histogram of the hyperfine field $\langle H_n \rangle$.
  (a) The spin distribution in the square unit cell
  assuming only the on-site hyperfine coupling $A = A_c = -23.8$
  T/$\mu_B$.
  (b) The spin distribution in the square unit cell
  assuming the transfer hyperfine coupling $B$ and $C$.
  On-site hyperfine coupling is $A = A_c - 4 B - C$.
  (c) The spin distribution in the rhomboid unit cell
  assuming only the on-site hyperfine coupling $A = A_c = -23.8$
  T/$\mu_B$.
  (d) The spin distribution in the rhomboid unit cell
  assuming the transfer hyperfine coupling $B$ and $C$.
  (e) The hyperfine field observed in NMR measurements.~\cite{kodama02b}
}
\end{figure}

The Cu NMR spectra was measured at $35$ mK in a field of $27.6$ T,
corresponding to the $1/8$-plateau. The overall shape of the 
spectra can be well reproduced assuming at least $11$ distinct
sites [see Fig.~\ref{fig:comp_exp}(e)]. If we consider only the on-site 
dominant hyperfine coupling, the hyperfine field is 
written as $H_n = A_c g_c \langle S_z \rangle$, where $\langle S_z
\rangle$ is the time-averaged local magnetization. The coupling 
constant $A_c$  and $g_c$-value are determined as $A_c = -23.8$
T/$\mu_B$ and $g_c = 2.28$ by electron spin resonance (ESR) and NMR 
measurements.~\cite{nojiri99,kodama02} 
Thus, the $11$ sites observed in NMR indicate
the existence of $11$ distinct spin sites in the $1/8$-plateau.
Two large positive $\langle S_z \rangle$ sites (two negative hyperfine
field sites) and one large negative $\langle S_z \rangle$ site
(one positive hyperfine field site) can be read off from
Fig.~\ref{fig:comp_exp}(e). In addition to them, several sites
spread around zero.

In this plateau, two types of unit cells 
(square and rhomboid cells) have been proposed theoretically
and so we calculated the spin textures in both unit cells as discussed
in Section~\ref{section:1/8-plateau}. In both cases,
there are two large positive spin sites and one large negative spin
site, which hardly depend on the shape of the unit cell.
However, the distribution of the expectation values 
of the spin component $\langle S_z \rangle$ on the other sites 
is different in the two cases.
In the square unit cell, the values $\langle S_z \rangle$
are concentrated around zero. On the other hand, in the
rhomboid one, the expectation values spread further away from zero. 
These facts indicate that
the results on rhomboid cell are qualitatively consistent with
the results of NMR. Hyperfine fields assuming
the on-site hyperfine coupling $A = A_c = - 23.8$ T/$\mu_B$ are shown
in Fig.~\ref{fig:comp_exp}(a) 
(Fig.~\ref{fig:comp_exp}(c)) for square (rhomboid) unit cell, in the
case of $\alpha = 10$.
Following Ref.~\onlinecite{kodama02b}, the hyperfine fields
including the effects of transfer hyperfine couplings $B$ and $C$ 
(see Fig.~\ref{fig:2d-model})
with the square (rhomboid) unit cell are also shown 
in Fig.~\ref{fig:comp_exp}(b) (Fig.~\ref{fig:comp_exp}(d)).
Here the parameters $B$ and $C$ are chosen to reproduce
the two large positive sites reported in the experiments. 
These results support the realization of the rhomboid cell.
However, it is difficult to estimate the realistic values of
the transfer hyperfine couplings and therefore the agreement of the
rhomboid cell with the experiment is only indicative.

It seems that the superstructure in the rhomboid unit cell  
is qualitatively consistent with the state observed by NMR.
There are still quantitative differences though.
For example, the numbers of different spin sites are 
not consistent: in the rhomboid cell, there are $8$ different spin
sites (see Fig.~\ref{fig:plateau0125b}) and in the experiments
at least $11$ sites exist. On the other hand, there are only $6$ different 
spin sites in the square unit cell. In the following we will discuss the 
possible effects of the inter-layer coupling and of the
Dzyaloshinsky-Moriya interaction on the local spin structure
and we emphasize the possible role of inter-layer coupling
to achieve a more quantitative agreement.

\subsection{inter-layer coupling}\label{section:il}

\begin{figure}
\vspace{2mm} 
\includegraphics[width=0.45\textwidth]{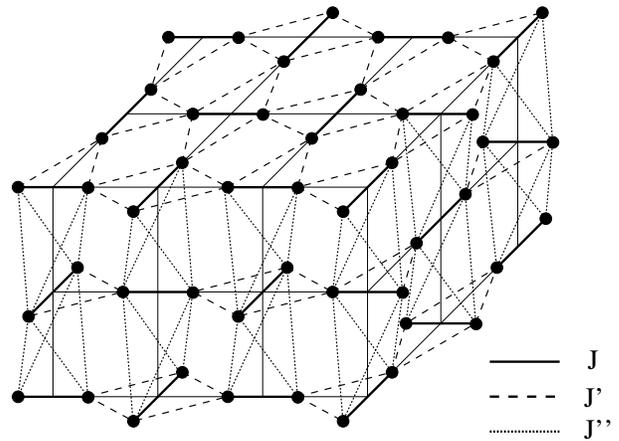}
\caption{\label{fig:3d-model}
The 3D orthogonal dimer model for ${\rm SrCu_2(BO_3)_2}$. 
}
\end{figure}

The three-dimensional (3D) structure of ${\rm SrCu_2(BO_3)_2}$ consists 
of ${\rm CuBO_3}$ layers, intercalated by magnetically 
inert ${\rm Sr}$ layers.
The magnetic ions ${\rm Cu^{2+}}$ form a three-dimensional
lattice structure shown in Fig.~\ref{fig:3d-model} and 
we denote the inter-layer coupling by $J^{\prime\prime}$.
However, we expect that antiferromagnetic coupling $J^{\prime\prime}$
is much smaller than the intra layer interactions $J$ and $J^{\prime}$
because of the presence of the ${\rm Sr}$ layers.
It is worth noting that, for small $J^{\prime\prime}/J$, the product of the 
dimer singlet, Eq.~(\ref{eq:dimer-singlet}), is the ground state also for the 
3D orthogonal dimer model.~\cite{ueda99,koga00b}
In addition, in this limit, the magnitude of the triplet excitations and 
their dispersion relation do not depend on the inter-layer interaction 
$J^{\prime\prime}$.~\cite{miyahara00b}
These facts further support the hypothesis that the magnetic properties of 
${\rm SrCu_2(BO_3)_2}$ can be very well described by the 2D 
orthogonal dimer model. However, the inter-layer interaction 
$J^{\prime\prime}$ may affect the magnetic properties
at high temperatures or under external magnetic 
fields. For instance, the inclusion of $J^{\prime\prime}$ is necessary 
to correctly reproduce the behavior of the magnetic susceptibility at high 
temperatures.~\cite{miyahara00b}

\begin{figure}
\vspace{2mm} 
\includegraphics[width=0.35\textwidth]{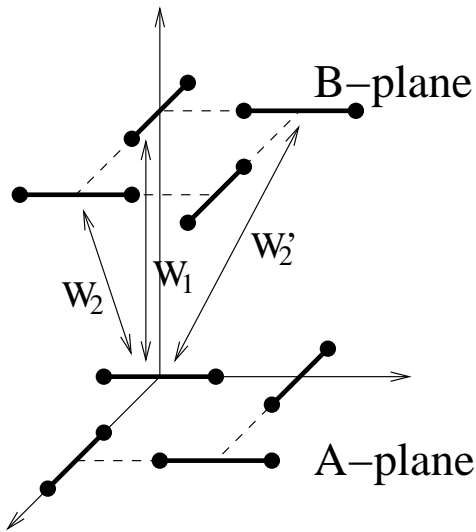}
\caption{\label{fig:interlayer_interactions}
Inter-layer interactions between the triplet excitations.
The nearest-neighbor interaction $W_1$, and the next-nearest-neighbor
interactions $W_2$ and $W_2^{\prime}$ are shown.
}
\end{figure}

Therefore, the effects of the inter-layer coupling might affect the 
superstructures at the plateaus. 
Although it is not presently possible to perform an exact calculation by
Lanczos diagonalization which includes a full 3D lattice, we try to combine 
the previous exact results with the perturbation theory based on the hard-core 
bosons to extract some conclusion on the possible 3D spin texture and on the 
possibility to obtain more than $8$ sites inside the unit cell.
Therefore, let us begin by considering the hard-core boson picture, and,
following Refs.~\onlinecite{momoi00,momoi00b,miyahara00}, we calculate the 
inter-layer interactions between triplets in two neighboring layers
(which we denote by A- and B-plane, see Fig.~\ref{fig:interlayer_interactions})
by using the perturbation theory in the limits $J^{\prime}/J \ll 1$ and
$J^{\prime\prime}/J \ll 1$.
Starting from a state where one of the singlets is promoted to a triplet on 
each layer, up to fifth order in $J^{\prime}/J$, the triplets are completely 
localized and the excitation energy is given by:
\begin{equation}
  E = 2 \Delta + W_k,
\end{equation}
where $\Delta$
is the spin gap energy for one triplet and
$W_k$ the interaction between the two triplets at distance $k$,
see Fig.~\ref{fig:interlayer_interactions}.
Up to third order, the spin gap energy
is written as $\Delta = J [1 - (J^{\prime}/J)^2 - 1/2 (J^{\prime}/J)^3]$
and the interaction energies are given by:
\begin{eqnarray}
  \frac{W_1}{J} & = & \frac{J^{\prime\prime}}{J}
  + \left( \frac{J^{\prime}}{J} \right)^2 \frac{J^{\prime\prime}}{J}, \\
  \frac{W_2}{J} & = & \frac{1}{2} \left( \frac{J^{\prime}}{J} \right)^2
  \frac{J^{\prime\prime}}{J}, \\
  \frac{W_2^{\prime}}{J} & = & 0.
\end{eqnarray}
The interaction energies vanish for $k>2$. 
Note that, similarly to what happens in the third-neighbor intra-layer 
interaction, the interactions for $k=2$ strongly depend on the relative
position of the triplets.~\cite{momoi00,momoi00b,miyahara00}

\begin{figure}
\vspace{2mm} 
\includegraphics[width=0.40\textwidth]{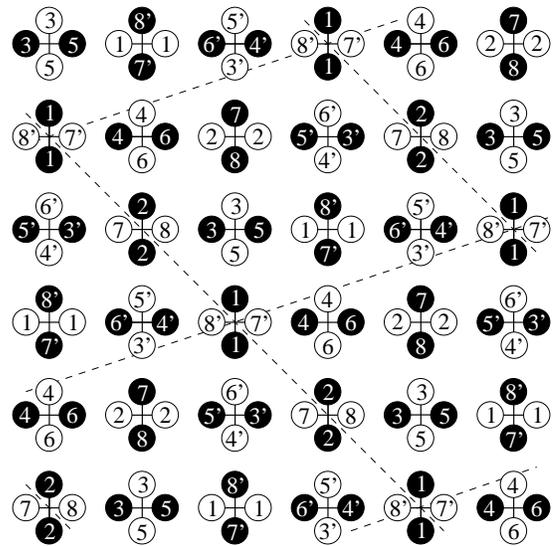}
\caption{\label{fig:spinzM0.125-B-il}
A stacking pattern for two neighboring layers. Black circles are in the A-plane
whereas white circles are in the B-plane. The numbers indicate the 
different spins within the unit cell (see Fig.~\ref{fig:plateau0125b}); 
the primed numbers are used to distinguish the spins modified by the 
inter-layer interaction.
}
\end{figure}

In the following, we consider the case of the rhomboid unit cell, which
seems to be a better condition to represent the real compound.
We remind that, in this
case, we have $8$ different sites, and therefore in each unit cell there are
two equivalent sites.
From the hard-core boson calculation, we can assume that the triplets on the 
B-layer are located as far as possible from the triplets on the A-layer.
In this assumption, the stacking pattern for the $1/8$-plateau is 
given in Fig.~\ref{fig:spinzM0.125-B-il}.
When the inter-layer coupling is considered, the two triplets of the cell 
repel each other and the two equivalent dimers in the cell of the A-layer 
face two {\it different} dimers in the cell of the B-layer.
Therefore, because of the inter-layer couplings, the two equivalent sites in a
given plane can split into two different local spins: Assuming the
stacking pattern of Fig.~\ref{fig:spinzM0.125-B-il}, we expect $14$ 
different spin states per unit cell.
In general, if we consider long-range interactions between two neighboring
planes, we can obtain up to $16$ different spins per unit cell.
However, the splitting for the sites indicated by $1$ and $2$ in
Fig.~\ref{fig:spinzM0.125-B-il} may be negligible because the difference 
originates from long-range interactions.

Therefore, from this analysis, it come out that the inclusion of the 
inter-layer coupling can give rise to $14-16$ different spin sites, which 
is in closer agreement with the NMR results, that indicate at least 
$11$ different sites per unit cell.
Note that in the square unit cell
the number of different spin sites remains $6$
even including the effect of inter-layer coupling $J^{\prime\prime}$.

\subsection{Dzyaloshinsky-Moriya interaction}\label{section:DM}

Finally, we consider the effect of the Dzyaloshinsky-Moriya interaction on
the spin texture.
Indeed, this term is relevant for ${\rm SrCu_2(BO_3)_2}$:
ESR and inelastic neutron scattering experiments show
an anisotropic behavior of the spin gap, which depends on
the direction of the external field~\cite{nojiri99,cepas01} and
such a behavior can be explained by considering
the Dzyaloshinsky-Moriya interaction.~\cite{cepas01,miyahara01}

If, following Ref.~\onlinecite{cepas01}, we assume in first approximation that 
the ${\rm CuBO_3}$ layer as the mirror plane, there is a
Dzyaloshinsky-Moriya term only on the $J^{\prime}$ bonds and 
its component is perpendicular to the plane.
In that case, the Hamiltonian for the Dzyaloshinsky-Moriya coupling is
\begin{equation}\label{eq:DM}
  {\cal H}_{\rm DM} = D_z
   \sum_{\rm (n.n.n.)} 
  (S_i^x  S_j^y - S_i^y S_j^x ).
\end{equation}
Note that Dzyaloshinsky-Moriya term ${\cal H}_{\rm DM}$ is odd under 
the exchange $i \leftrightarrow j$ and therefore, we have to fix 
the direction from $i$ to $j$ for a pair $i, j$, as shown in
Fig.~\ref{fig:2d-model}.

\begin{figure}
\vspace{2mm} 
\includegraphics[width=0.45\textwidth]{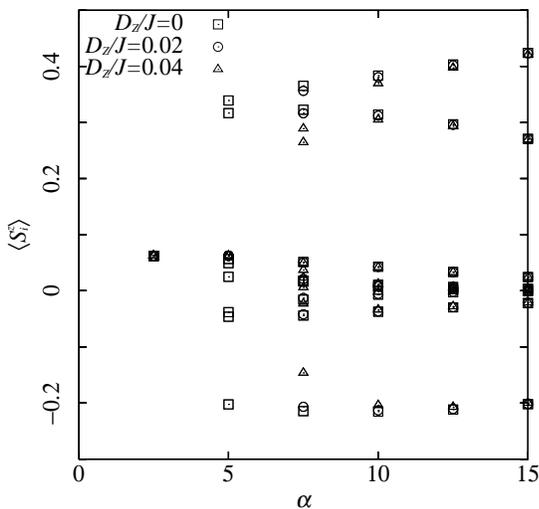}
\caption{\label{fig:DM}
Local magnetization for the eight different sites in the rhomboid unit
cell for the $1/8$-plateau as a function of the spin-phonon coupling
for different values of the Dzyaloshinsky-Moriya interaction $D_z$.
}
\end{figure}

Strictly speaking, in the real compound, there is a slight buckling of the 
${\rm CuBO_3}$ plane~\cite{sparta01} and so other higher components
of Dzyaloshinsky-Moriya interaction can appear.
However, the magnitude of this buckling is very small and therefore
the higher terms might be neglected with respect to the one given by 
Eq.~\ref{eq:DM}.
In practice, we have included the Dzyaloshinsky-Moriya interaction, 
Eq.~(\ref{eq:DM}), in the original Hamiltonian~(\ref{hamilt}) and we
have performed our self-consistent Lanczos diagonalization of the 2D cluster.
The results for the rhomboid cluster with 
$D_z/J = 0.02$ and $0.04$ are shown 
in Fig.~\ref{fig:DM} and compared with the ones for $D_z=0$. 
The inclusion of the Dzyaloshinsky-Moriya 
interaction makes the uniform state more stable for small spin-phonon 
coupling, and the inhomogeneous state appears for a spin-phonon coupling 
larger that the one we obtain without this interaction. 
The reason for this shift is due to the fact that the Dzyaloshinsky-Moriya 
term favors the hopping of the triplet.
On the other hand, for strong enough spin-phonon couplings, i.e., 
$\alpha \gtrsim 7$, the values of the local spins are only slightly modified
by the Dzyaloshinsky-Moriya interaction, and the number of different spins
is always equal to $8$.

Therefore, we can infer that the Dzyaloshinsky-Moriya interaction is not
a fundamental ingredient in determining the spin texture at the magnetization
plateau, and, in a first approximation, can be neglected.

\section{Conclusion}\label{section:conclusions}

In this paper, we have considered the effect of adiabatic phonons on the 2D 
orthogonal dimer model by using  a self-consistent Lanczos diagonalization
of small clusters. This study is directly related to the properties of
${\rm SrCu_2(BO_3)_2}$, a new spin gapped material, under magnetic field.
From an experimental point of view, at present, the only available informations
about the magnetic texture are about the $1/8$-plateau, which is 
stabilized for magnetic fields around $27.6$ Tesla.
Our theoretical results for this plateau indicate
that the spin-phonon coupling 
is able to stabilize two possible candidates for the local spin texture: 
i) a square unit cell of $16$ sites with $6$ different values 
of the spins and ii) a rhomboid unit cell of $16$ sites with $8$ different
values of the spins. Although the energy difference between the two
configurations is very small, the rhomboid cell contains a larger number
of different spins and therefore it describes better the 
experimental finding.
A closer agreement with the NMR results can be achieved by the inclusion of
the inter-layer coupling, that can split equivalent sites, whereas the
Dzyaloshinsky-Moriya interaction does not seem to play an important role
in determining the actual values of the local spins.
From our calculations, it comes out that the superstructure
at the $1/8$-plateau 
is qualitatively consistent with the NMR results, but it is still open 
question to make a more quantitative comparison.
The main limitation of our Lanczos calculations is the
presence of finite size effects. As we already emphasized previously, 
the finite size effects might be small because the decay of the Friedel-like 
oscillation of the local spins is quite fast.
However, in a small cluster, the uniform state is easily stabilized, and,
therefore, the spin-phonon coupling required might be bigger than the 
realistic value (in the thermodynamic limit). Second,
as mentioned in Section~\ref{section:comp}, 
the inclusion of the transferred hyperfine couplings 
to neighboring spins is probably needed to reach a better agreement.
Indeed we showed that the inclusion of the 
transferred hyperfine coupling to the nearest- and 
next-nearest-neighbors improves the agreement between experiments and 
theoretical results. 
Moreover, it might be necessary to include 
also transferred hyperfine couplings from sites on neighboring
layers. Assuming the stacking pattern in Fig.~\ref{fig:spinzM0.125-B-il},
such hyperfine couplings can induce different hyperfine
fields at two equivalent sites of a given plane.
Thus these hyperfine couplings enhance the effects of the inter-layer
couplings which we discussed in Section~\ref{section:il}. 
However, it is difficult to obtain reliable estimates of
the transferred hyperfine couplings, and this discussion can 
be only indicative. 

Although at present, there are no experimental results on the higher
plateaus, we studied also the effects of the spin-phonon coupling for
$1/4$-, $1/3$-, and $1/2$-plateaus, with
the hope that in the near future it will
be possible to have experimental insight into the local magnetization of 
these plateaus (or at least some of them).
Our results indicate that stripe-like superstructures
are realized at the
$1/3$- and $1/4$-plateaus, whereas a structure with a square unit cell is
stabilized at the $1/2$-plateau. 
It is worth noting that all the results regarding the shape of the unit 
cells are consistent with the hard-core boson picture, 
although within this hard-core boson approximation
it is not possible to obtain more than two different sites.
In this respect, our Lanczos results give valuable insight into the real 
magnetization pattern that appears in the fascinating regions of the 
magnetization plateaus.

\acknowledgments
It is a pleasure to thank K. Kodama, M. Takigawa, M. Horvati\'c, 
and C. Berthier
for showing and explaining to us their experimental results
before publication. We also thank K. Ueda, H. Kageyama, 
D. Poilblanc, and T. Ziman for useful discussion.
This work has been supported by the Swiss National Fund.

%%%%%%%%%%%%%%%%%%%%%%%%%%%%%%%%%%%%%%%%%%%%%%%%%%%%%%%%%%%%%%%%%%%%%%%%
%                           BIBLIOGRAPHY
%%%%%%%%%%%%%%%%%%%%%%%%%%%%%%%%%%%%%%%%%%%%%%%%%%%%%%%%%%%%%%%%%%%%%%%%

\end{document}